\documentclass[12pt]{article} 

\errorstopmode

\usepackage{graphicx}
\usepackage{amssymb}
\usepackage{amsmath}

\newcommand{\dd}{{\rm d}}
\newcommand{\ii}{{\rm i}}
\newcommand{\fii}{\varphi}
\newcommand{\e}{{\rm e}}
\DeclareMathOperator*{\Ai}{Ai}
\DeclareMathOperator*{\Bi}{Bi}

\begin{document}
\title{\bf Spectra of absolute instruments \\ from the WKB approximation}

\author{Tom\'a\v{s} Tyc \\
Faculty of Science and Faculty of Informatics, Masaryk University,\\
Kotl\'a\v rsk\'a 2, 61137 Brno, Czech Republic \\} 
\date{\today}
\maketitle
\begin{abstract}

We calculate frequency spectra of absolute optical instruments using the WKB
approximation. The resulting eigenfrequencies approximate the actual values
very accurately, in some cases they even give the exact values.  Our
calculations confirm results obtained previously by a completely different
method.  In particular, the eigenfrequencies of absolute instruments form tight
groups that are almost equidistantly spaced. We demonstrate our method and its
results on several examples. \\[3mm]

% PACS: 42.30.Va

\end{abstract}

\section{Introduction}

Absolute optical instrument (AI) is a device that provides a perfectly sharp
image of all points in some spatial region~\cite{BornWolf}.  The simplest AI is
a plane mirror that gives a virtual image of a whole half-space. Another,
beautiful example of an AI is Maxwell's fish eye, discovered by J.~C.~Maxwell
in 1854~\cite{MFE}, that images sharply the whole space, and all rays form
circles. In recent years absolute instruments attracted an increased interest
which has led to proposing new devices of various types, e.g. AIs that perform
imaging of the whole space, of optically homogeneous
regions~\cite{minano06,Tyc11njp}, or provide magnified
images~\cite{Tyc11pra}. A general method has been proposed in~\cite{Tyc11njp}
for designing spherically symmetric AIs. This research was based on geometrical
optics.

Recently AIs attracted attention also from the point of view of wave optics. It
was shown both theoretically~\cite{Ulf2009-fisheye} and
experimentally~\cite{Ulf2011,Ulf2012} that these devices can provide
subwavelength resolution, although this claim has raised controversy
\cite{Blaikie,Ulf-reply_to_Blaikie,Tyc2011-nature} and it is still not clear to
what extent such a super-resolution can be used practically~\cite{Rhiannon}. A
different question was addressed in ~\cite{Tyc12njp}, namely what are the
general characteristics of the spectrum of eigenfrequencies of AIs. It was
shown that the spectrum consists of tight groups of levels with almost
equidistant spacing between them. This finding was based on an analysis of a
light pulse propagating in the AI and on the assumption that a short pulse
emitted at some point can be absorbed at the image point during a short time as
well. Numerically calculated spectra of various AI confirmed this theoretical
result very well.

In this paper, we investigate the spectra of absolute instruments by a
completely different method. Employing the WKB approximation with the Langer
modification and using one of the general properties of AIs, we confirm in a
different way the previously known results about their
spectrum~\cite{Tyc12njp}. Our method has two advantages compared to the
previous one: it enables to calculate not just the spacing of the level groups
but also their offset, and it allows to treat the situations where a mirror is
used in the device. We verify our results by comparing the calculated spectra
with numerical values for several examples of AIs.

The paper is organised as follows. In Sec.~\ref{absolute} we recall absolute
instruments and discuss some of their properties. In Sec.~\ref{wkb} we employ
the WKB method for calculating the spectra of radially symmetric media and in
Sec.~\ref{examples} we illustrate the results on particular examples of AIs.
In Sec.~\ref{mirror} we analyse the situation in AIs that contain mirrors, and
we conclude in Sec.~\ref{conclusions}.

\section{Absolute optical instruments}
\label{absolute}

In this section we recall some properties of absolute instruments from the
point of view of geometrical optics, which will be useful for the subsequent
calculations. We will consider radially symmetric AIs with the refractive index
distribution $n(r)$. In addition, we will focus on a specific class of absolute
instruments, namely AIs of the first type~\cite{Tyc11njp}.  AI of the first
type is a device with the property that every point A from its whole volume has
a full image at some point B, which means that all rays emerging from A reach
B. Since the role of the points A and B can be interchanged, it is clear that
any ray emerging from the point A returns back to this point again, so A is an
image of itself. In some AIs, it is the only image, in other ones there may be
more images.

The AI can be either three- or two-dimensional. In the latter case we consider
a 2D propagation of rays in a 2D refractive index profile. Due to the radial
symmetry of the device, the following quantity analogous to the mechanical
angular momentum is conserved~\cite{BornWolf}:
\begin{equation}
   {L}=rn(r)\sin\alpha\,,
\label{L}\end{equation}
where $\alpha$ is the angle between the tangent to the particle trajectory and
the radius vector.  In 3D, another consequence of the radial symmetry of the
device is that each ray lies in a plane containing the centre of symmetry of
the device O, so the ray trajectory can be described in polar coordinates
$(r,\fii)$ both in the 2D and 3D cases.  For nonzero ${L}$, the polar angle
$\fii$ increases (or decreases) monotonically while $r$ oscillates between the
turning points $r_-$ and $r_+$, $r_-\le r_+$.  We will assume that for each
possible value of $L$, there are just two turning points that coalesce when $L$
reaches its largest possible value $L_0$.

In order to ensure that any ray emerging from a point A eventually returns
there, the change of the polar angle $\Delta\fii$ corresponding to $r$ changing
from $r_-$ to $r_+$ must be a rational multiple of $2\pi$, so we can write
$\Delta\fii=\pi/\mu$, $\mu\in{\mathbb Q}$\footnote{In~\cite{Tyc11njp}, $m$ was
  used instead of $\mu$.}. We will assume in this paper that the value of $\mu$
is the same for all possible angular momenta $L$, which is the most usual
case. With respect to this assumption and since any point is an image of
itself, the optical path $S_{\rm AA}=\int n\,\dd l$ from A back to A is equal
for all rays~\cite{BornWolf}.  At the same time, the ray from A back to A
consists of an integer number of segments on each of which $r$ changes from
$r_-$ to $r_+$ or back.  Therefore also the optical path $S$ between two turning
points has to be equal for all rays, i.e., it is independent of the angular
momentum $L$. This optical path can in general be expresses as
\begin{equation}
  S({L})=\int_{r_-}^{r_+}\frac{n\,\dd r}{\cos\alpha}=\int_{r_-}^{r_+}
   \frac{n\,\dd r}{\sqrt{1-({L}/nr)^2}}\,,
\label{S}
\end{equation} 
where we have used Eq.~(\ref{L}) and written the dependence on $L$
explicitly. The fact that for absolute instruments $S$ is independent of $L$ is
of a key importance for calculation of the spectra with the WKB method, as we
will see in the following section.

As the last thing we will express the optical path between the turning points
in a different way. When $L$ increases, the turning points $r_\pm$ approach
each other and finally meet at the point $r_0$, the radius of the circular ray,
for the maximum possible value of angular momentum $L=L_0$. The corresponding
optical path $S$ is then equal to the geometrical path $\Delta\fii r_0=\pi
r_0/\mu$ multiplied by the refractive index $n(r_0)$, i.e.,
\begin{equation}
  S=\frac{\pi r_0n(r_0)}\mu \,.
\label{S0}
\end{equation}

\section{WKB calculation of spectra of AIs}
\label{wkb}

For simplicity we will consider a monochromatic scalar wave in the AI that can
be described by the Helmholtz equation
\begin{equation}
  \Delta\psi+k^2 n(r)^2 \psi=0\,.
\label{}
\end{equation} 
If the speed of light is set to unity, $k$ is at the same time equal to the
frequency $\omega$ of the wave.  Separating the radial and angular parts in
terms of $\psi(r,\theta,\fii)=R(r)Y_{lm}(\theta,\fii)$ (in 3D) or
$\psi(r,\fii)=R(r)\e^{\ii m\fii}$ (in 2D) and making the substitution
$R(r)=r^{-1}w(r)$ (in 3D) or $R(r)=r^{-1/2}w(r)$ (in 2D), we get the equation
for $w(r)$ in the form
\begin{equation}
  w''+\left[-\frac{l(l+1)}{r^2}+k^2 n^2\right]w=0\quad\mbox{(3D)},\qquad
  w''+\left[-\frac{m^2-1/4}{r^2}+k^2 n^2\right]w=0\quad\mbox{(2D)}\,,
\label{for_w}
\end{equation} 
where the prime denotes a derivative with respect to $r$. 

The equations~(\ref{for_w}) are analogous to the radial part of the
Schr\"odinger equation obtained when solving a quantum-mechanical motion in a
central potential.  As was shown by Langer~\cite{Langer37}, however, a direct
application of the WKB method to that equation leads to problems due to the
centrifugal potential and termination of the $r$-axis at $r=0$, and yields
wrong results for the energy spectrum of the system. A careful analysis shows
that the problem can be eliminated by the substitution $l(l+1) \to (l+1/2)^2$
(or $m^2-1/4\to m^2$ in 2D) --- the so-called Langer modification,
see~\cite{Langer37,Berry72} or \S~49 of~\cite{Landau}. Then the WKB method can
be applied to the resulting equation directly and gives correct
results. Exactly the same argument applies also in our case, which leads to the
equations
\begin{equation}
  w''+\left[-\frac{(l+1/2)^2}{r^2}+k^2 n^2\right]w=0\quad\mbox{(3D)}\,,
\label{for_w2}
\end{equation} 
and 
\begin{equation}
  w''+\left[-\frac{m^2}{r^2}+k^2 n^2\right]w=0\quad\mbox{(2D)}\,.
\label{for_w2-2D}
\end{equation} 
We can now apply the WKB approximation to these equations directly.  In the
following we will write the formulas just for the 3D case as the treatment of
the 2D case is analogous and can be obtained from the 3D results by the
substitution $l+1/2\to m$.

We first substitute $w=\rho\e^{\ii\phi}$ in Eq.~(\ref{for_w2}), which yields
\begin{equation}
  \rho''-\rho\phi'^2+\ii(2\rho'\phi'+\rho\phi'')
  +\left[k^2n^2-\frac{(l+1/2)^2}{r^2}\right]\rho=0
\label{}
\end{equation} 
Then we separate the real and imaginary parts and in the real part neglect the
term $\rho''$ with respect to $\rho\phi'^2$; this corresponds to the
first order of WKB approximation. This way we get two equations
\begin{equation}
 \phi'=\pm p(r)\,,\qquad  \rho\phi''+2\rho'\phi'=0\,, 
\label{eq-wkb}
\end{equation} 
where we have denoted
\begin{equation}
 p(r)=\sqrt{k^2n^2-\frac{(l+1/2)^2}{r^2}}\,.
\label{p}
\end{equation} 
The first equation in (\ref{eq-wkb}) has the solution $\phi(r)=\pm\int
p(r)\,\dd r$, the second one has the solution $\rho=\mbox{const}\times
p^{-1/2}$. Combining these solutions, we get 
\begin{equation}
 w(r)=\frac{\mbox{const}}{\sqrt{|p|}}\exp\left(\pm \ii\int p(r)\,\dd r\right).
\label{w}
\end{equation} 
The function $w(r)$ has an oscillatory behaviour in the region where $p$ is
real, i.e., between the turning points $r_\pm$ at which $p$ turns to zero. For
$r<r_-$ and $r>r_+$, $p$ is imaginary and the function $w(r)$ describes
evanescent waves in these regions.  We now need to match the oscillatory and
evanescent solutions in the three regions, which will give the quantisation
condition for $k$. This matching, however, cannot be done within the WKB method
itself because the condition $|\rho''|\ll|\rho\phi'^2|$ breaks up at the
turning points. Still there are several ways how to get around this
problem~\cite{Landau}. One of them is to treat $r$ formally as a complex
variable and circumvent the turning point in the complex $r$-plane. Another
standard method is to approximate the bracket in Eq.~(\ref{for_w2}), i.e.,
$p^2$, by a linear function of $r$ in the vicinity of the turning point (say
$r_+$). Both of these methods give the same well-known result~\cite{Landau}:
the wave decaying for $r>r_+$ matches the following wave in the oscillatory
region:
\begin{equation}
  \frac{\mbox{const}}{\sqrt{p}}\cos\left(\int_{r_+}^r p(r')\,\dd r'
               +\frac\pi4\right) .
\label{approxcos}
\end{equation} 
This will be discussed in detail for a more general case in Sec.~\ref{mirror}.
A similar consideration can be made for the other turning point $r_-$. To
satisfy both the phase conditions simultaneously, it must hold
\begin{equation}
 \Phi(l,k)=\pi\left(N+\frac12\right)\,,\quad N=0,1,2,\dots\,,
\label{cond}
\end{equation} 
where we have defined
\begin{equation}
   \Phi(l,k)\equiv \int_{r_-}^{r_+} p(r)\,\dd r
    =\int_{r_-}^{r_+}\sqrt{k^2n^2-\frac{(l+1/2)^2}{r^2}}\,\dd r\,.
\label{Phi}
\end{equation} 
Eq.~(\ref{cond}) is equivalent to the Born-Sommerfeld quantisation rule in
quantum mechanics~\cite{Landau} and together with Eq.~(\ref{Phi}) it presents
the main result of this section.
 
In some situations the condition imposed on the function $w(r)$ may be
different.  In particular, if there is a mirror in the AI at some radius $r=b$,
then the wave must vanish there. This condition will then lead to a different
phase shift in Eq.~(\ref{approxcos}) and hence to a shift in the
eigenfrequencies, which will be discussed in Sec.~\ref{mirror}.

\subsection{Calculation of the integral in  Eq.~(\ref{cond})}
\label{integral}

To find the eigenfrequencies for which the condition~(\ref{cond}) is satisfied,
we need to evaluate the integral~(\ref{Phi}). Quite remarkably, $\Phi(l,k)$ is
closely related to the optical path~$S$ in the $(r,\fii)$ plane between the
turning points, defined by Eq.~(\ref{S}). It is easy to check that
\begin{equation}
  \frac{\partial\Phi(l,k)}{\partial k}=S\left(\frac{l+1/2}k\right).
\label{der}
\end{equation} 
As we have seen in Sec.~\ref{absolute}, in AIs $S(L)$ is independent of $L$,
and therefore the right-hand side of Eq.~(\ref{der}) does not depend on $k$ for
a fixed $l$. This allows us to integrate Eq.~(\ref{der}) to get
\begin{equation}
 \Phi(l,k)=Sk+c(l)\,,
\label{sol}
\end{equation} 
where $c(l)$ is an integration constant with respect to $k$. To find it, we
consider the situation when the turning points $r_\pm$ (given by the condition
$p=0$) coincide. As we have seen in Sec.~\ref{absolute}, this corresponds to
the value $L=r_0n(r_0)$ in Eq.~(\ref{S}). Comparing the square roots in
Eqs.~(\ref{S}) and~(\ref{p}), we see that the corresponding value of $k$ is
$k_0=(l+1/2)/[r_0n(r_0)]$. Moreover, in this situation $\Phi=0$ because
$r_+=r_-$.  Eq.~(\ref{sol}) then yields the constant
$c(l)=-(l+1/2)S/(r_0n(r_0))$. Substituting this to Eq.~(\ref{sol}) and using
Eq.~(\ref{S0}), we get for $k$
\begin{equation}
   k=\frac1{r_0n(r_0)}\,\left[\frac{\mu\Phi}\pi+\left(l+\frac12\right)\right]
\label{kphi}
\end{equation} 
Combining Eqs.~(\ref{kphi}) and~(\ref{cond}), we get finally the
quantisation condition for the eigenfrequencies of a 3D absolute instrument
\begin{equation}
  k=\frac1{r_0n(r_0)}\,\left[\mu\left(N+\frac12\right)+l+\frac12\right]\,\quad
  N=0,1,2,\dots\,
\label{k3D}
\end{equation} 
The spectrum of a 2D AI is obtained by replacing  $l+1/2$ by $|m|$:
\begin{equation}
  k=\frac1{r_0n(r_0)}\,\left[\mu\left(N+\frac12\right)+|m|\right]\,\quad
  N=0,1,2,\dots 
\label{k2D}
\end{equation} 
The formulas~(\ref{k3D}),~(\ref{k2D}) for the WKB-approximated spectrum of
absolute instruments form the central result of this paper. They are based on
the key relation~(\ref{der}) and on the fact that in AIs the optical path
$S$ is independent of the angular momentum.

Inspection of the formulas~(\ref{k3D}),~(\ref{k2D}) shows two things. First,
for absolute instruments the WKB spectrum is degenerate, and this degeneracy
increases with the frequency. This is obvious from the fact that $\mu$ is a
rational number for AIs, and usually it is a ratio of small integers. Then we
can find a number of combinations $(l,N)$ yielding the same frequency $k$, and
this number increases with increasing $l$ and/or $N$. Since the
formulas~(\ref{k3D}) and~(\ref{k2D}) give the WKB approximation to the
spectrum, the exact spectrum exhibits this degeneracy only approximately and
the eigenfrequencies form tight groups.  The second property that follows from
Eqs.~(\ref{k3D}) and~(\ref{k2D}) is that the frequency groups are positioned
equidistantly because $\mu$ is rational and $m,l$ and $N$ all change in integer
steps.  As was shown in~\cite{Tyc11njp}, both of these properties of the
spectrum have key importance for the ability of AIs to focus waves.
Eqs.~(\ref{k3D}) and~(\ref{k2D}) not only confirm the previous results, but in
addition enable to calculate the absolute position of the spectral structure,
which was not possible by the previous method.

\section{Examples}
\label{examples}

To see how well the WKB spectrum approximates the actual spectrum, let us now
apply our results on several examples of AIs. We will represent the spectra
graphically by the function $k(\nu)$ that has been introduced
in~\cite{Tyc12njp} (the notation $\omega(\nu)$ was used instead). Its value at
integer $\nu$ is simply the $\nu^{\rm th}$ eigenfrequency and for the
non-integer $\nu$, $k(\nu)$ is given by linear
interpolation~\cite{Tyc12njp}. Degenerate levels are clearly exhibited in the
graph of $k(\nu)$ as intervals where the function is constant.

\begin{figure}
\begin{center}
\begin{tabular}{ccc}
\includegraphics[width=8cm]{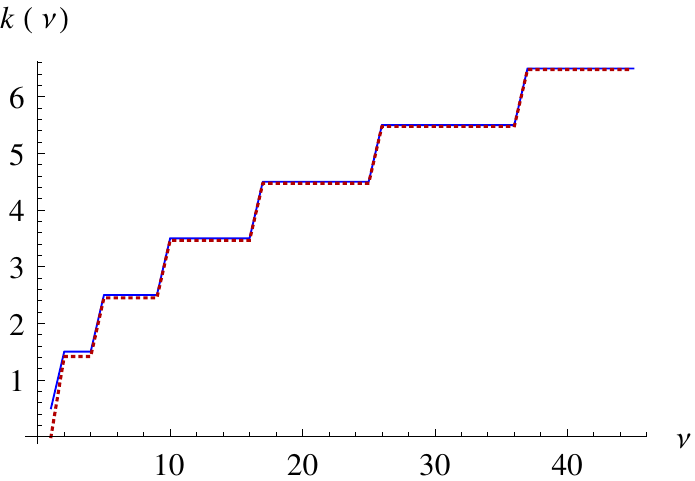}&\hspace{5mm}&
\includegraphics[width=8cm]{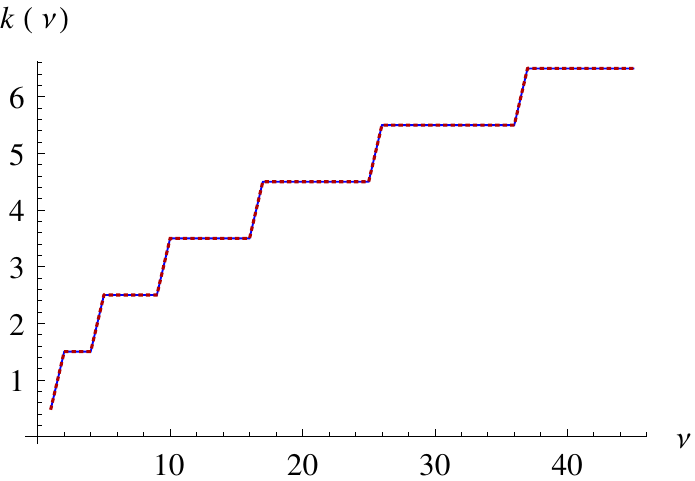}\\
(a)&&(b)\\
\includegraphics[width=8cm]{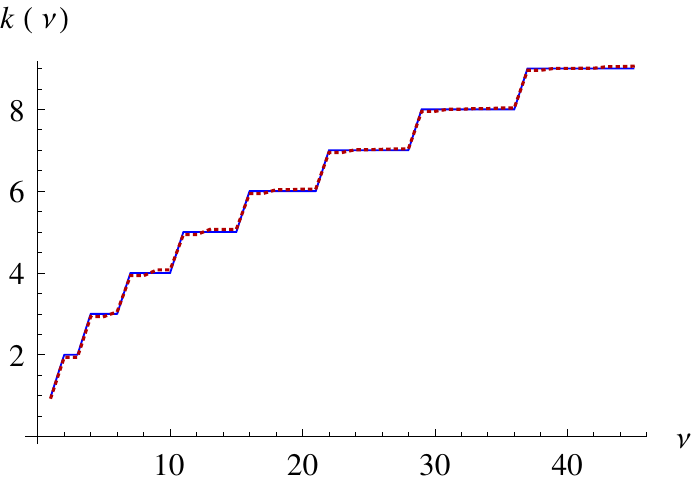}&\hspace{5mm}&
\includegraphics[width=8cm]{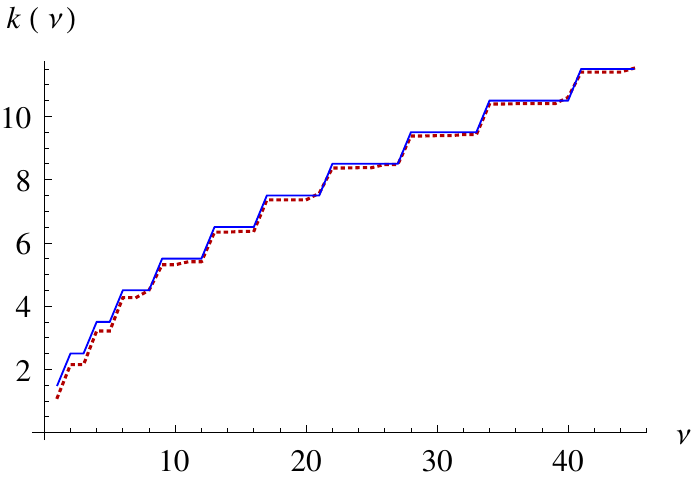}\\
(c)&&(d)
\end{tabular}
\end{center}
\caption{Comparison of the spectra of several AIs calculated with the WKB
  method (blue line) with the analytically or numerically calculated spectra
  (red dashed line). The spectra are represented by the function $k(\nu)$,
  degenerate levels correspond to plateaus in the graph. (a) Maxwell's fish
  eye, (b) Kepler profile, (c) Mi\~nano lens, (d) a 2D AI with the refractive
  index $n(r)=2r^{-1}\{[(2-r)/r]^{3/2}+[r/(2-r)]^{3/2}\}^{-1}$. In (b), the
  analytical and WKB spectra are identical. }
\label{fig-spectra}
\end{figure}

\begin{itemize}
\item {\bf Maxwell's fish eye}, $n(r)=2/(1+r^2)$. In this case $\mu=1$,
  $r_0=1$, which yields the WKB spectrum in 2D
\begin{equation}
   k_{\rm WKB}=N+|m|+1/2\,.
\label{kmfe1}
\end{equation} 
The analytic form of the spectrum is
\begin{equation}
   k_{\rm analytic}=\sqrt{(N+|m|)(N+|m|+1)}\,.
\label{kmfe2} 
\end{equation} 
With the exception of the lowest level, Eq.~(\ref{kmfe1}) approximates the
correct values~(\ref{kmfe2}) very well: for $N+|m|=1$ the relative error is
only $6\%$, for $N+|m|=3$ it is $1\%$, and it further decreases with increasing
$N+|m|$. The two spectra are compared graphically in
Fig.~\ref{fig-spectra}~(a). For the 3D Maxwell's fish eye profile, the WKB and
exact eigenfrequencies are given by the same equations~(\ref{kmfe1})
and~(\ref{kmfe2}), respectively, with $|m|$ replaced by $l+1/2$. Again the
agreement is very good.
\item {\bf Hooke index profile} with the refractive index $n(r)=\sqrt{2-r^2}$.
  This formula for the index is used also for $r>\sqrt2$ where $n$ becomes
  imaginary. This allows to calculate the eigenmodes and eigenfrequencies
  analytically as well; they are given by the Laguerre-Gaussian functions
  similarly as stationary states of an isotropic 2D harmonic oscillator in
  quantum mechanics. In this case $\mu=2$, $r_0=1$, which yields the 2D WKB
  spectrum
 \begin{equation}
   k_{\rm WKB}=2N+|m|+1\,.
\label{kHooke}
\end{equation} 
This is at the same time also the analytical form of the spectrum, so the WKB
approximation yields the exact values for the eigenfrequencies in this
case. This is in a complete analogy with the situation in quantum mechanics
where the WKB approximation also yields the exact values for the energy
eigenvalues of a 2D harmonic oscillator. For the 3D Hooke profile both spectra
are given by Eq.~(\ref{kHooke}), again with $|m|$ replaced by $l+1/2$; the
agreement is perfect again.

\item {\bf Kepler index profile}, $n(r)=\sqrt{2/r-1}$. In this case $\mu=1$,
  $r_0=1$, which yields the 2D WKB spectrum in the form~(\ref{kmfe1}). This is
  at the same time also the analytical spectrum (again the formula for $n(r)$
  is used also for $r>2$), so again the WKB gives here the
  exact spectrum.  The spectra are shown in Fig.~\ref{fig-spectra}~(b). The
  situation in 3D is analogous.

\item {\bf 2D Mi\~nano lens}, $n(r)=1$ for $r\le1$ and $n(r)=\sqrt{2/r-1}$ for
  $r\ge1$. In this case $\mu=2$, $r_0=1$, which yields  the WKB spectrum in the
  form~(\ref{kHooke}). Comparison with the numerically calculated
  spectrum is shown in Fig.~\ref{fig-spectra}~(c). The agreement is
  very good.
\item As the last example we consider a 2D AI with the refractive index
  $n(r)=2r^{-1}\{[(2-r)/r]^{3/2}+[r/(2-r)]^{3/2}\}^{-1}$ derived from the
  general formula (18) of~\cite{Tyc11njp} with $f(r)=2-r$ and $\mu=3$. This
  yields for the WKB spectrum
 \begin{equation}
   k_{\rm WKB}=3N+|m|+\frac32\,,
\label{km=3}
\end{equation} 
which again approximates the exact spectrum well, as can be seen in
Fig.~\ref{fig-spectra}~(d).
\end{itemize}
We see that in all these cases the agreement of the WKB spectrum with the exact
one is very good, even for the lowest states. This is quite a remarkable
feature of the WKB approximation that our optical situation shares with quantum
mechanics.

\section{Mirrors}
\label{mirror}

So far we have considered absolute instruments without mirrors, so we required
the wave in the evanescent region to decay gradually. However, in many AIs
there are mirrors that constitute their important parts. The mirrors play one
or both of the following two roles. First, in some AIs they limit the size of
the device, e.g. from infinite size to a finite one. This is the case of
Maxwell's fish eye mirror (MFEM)~\cite{Ulf2009-fisheye,Tyc11njp}. The second
possible role is that mirrors can eliminate regions where the refractive index
goes to zero; this is again the case of MFEM or e.g. of the modified Mi\~nano
lens~\cite{Tyc11njp}.

Consider a situation that there is a spherical (or in 2D, circular) mirror at
radius $r=b$ in the AI. Due to the boundary condition the wave has to vanish at
it, i.e., $w(b)=0$. We have to distinguish two cases: either the mirror is
placed in the evanescent or in the oscillatory region. In the following we will
treat them separately.

\subsection{Mirror in the evanescent region}

Suppose first that the mirror is placed in the evanescent region,
e.g. $b>r_+$. Then for $r>r_+$ we need not only the exponentially decaying
solution in Eq.~(\ref{w}) but also the exponentially growing one to create
their superposition that vanishes at $r=b$.  This will cause a phase shift of
the wave in the oscillatory region and consequently a shift in the spectrum
compared to the situation without the mirror.

\begin{figure}
\begin{center}
\begin{tabular}{ccccc}
\includegraphics[width=5cm]{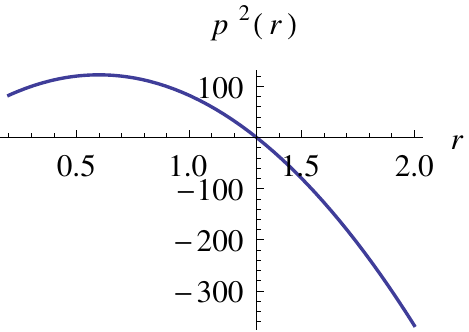}&\hspace{5mm}&
\includegraphics[width=5cm]{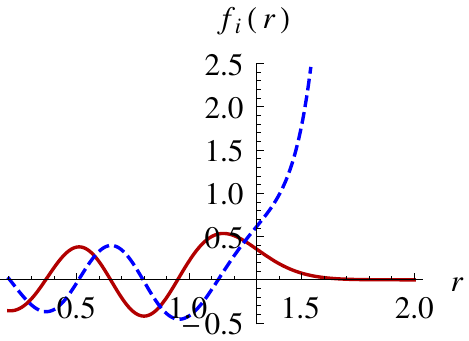}&\hspace{5mm}&
\includegraphics[width=5cm]{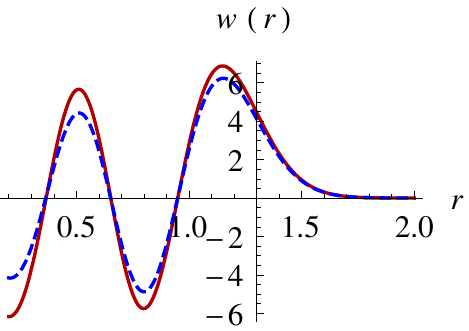}\\
(a)&&(b)&&(c)
\end{tabular}
\end{center}
\caption{Example of the functions relevant for describing the approximate
  solutions of Eq.~(\ref{for_w2}). (a) The function $p^2(r)$ chosen to be
  significantly non-linear, (b) the corresponding functions $f_1(r)$ (red) and
  $f_2(r)$ (dashed blue) from Eqs.~(\ref{Airysol}). In (c) we compare a 
  numerical
  solution of Eq.~(\ref{for_w2}) satisfying $w(2)=0,w'(2)=-0.02$ (red curve)
  with its approximate solution in terms of a superposition of the functions
  $f_1(r),f_2(r)$ (dashed blue curve) satisfying the same conditions. The
  agreement between the two curves is very good, in particular the position of
  the nodes that has the key importance for the spectrum. The $y$-axis of the
  graphs is placed at the turning point $r_+$ (in this case $r_+=1.3$). }
\label{figure-airyfun}
\end{figure}

\begin{figure}
\begin{center}
\begin{tabular}{ccccc}
\includegraphics[width=5cm]{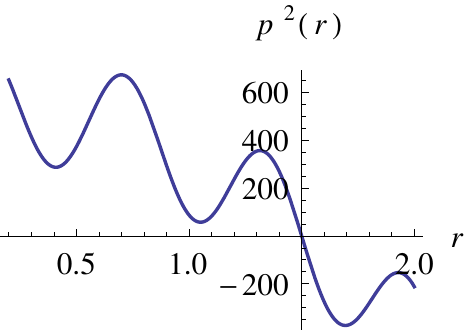}&\hspace{5mm}&
\includegraphics[width=5cm]{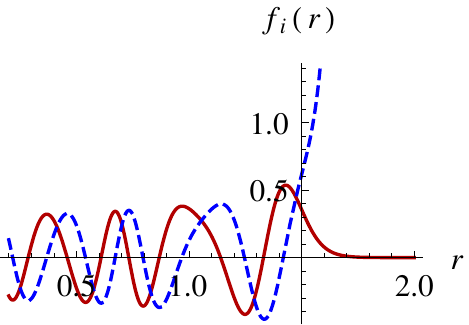}&\hspace{5mm}&
\includegraphics[width=5cm]{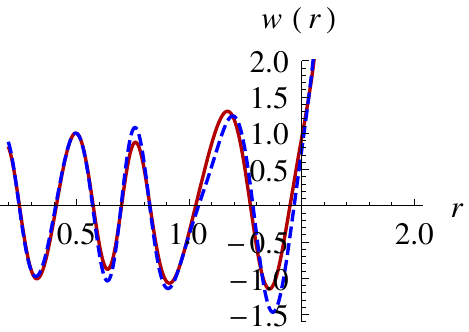}\\
(a)&&(b)&&(c)
\end{tabular}
\end{center}
\caption{As in Fig.~\ref{figure-airyfun}, but with an even stronger
  non-linearity of the function $p(r)^2$, and the parameters $r_+=1.5$,
  $w(0.5)=1,w'(0.5)=-1$. }
\label{figure-airyfun2}
\end{figure}

To describe such a situation, we will use one of the standard methods for
matching the WKB solutions in the oscillatory and evanescent regions employing
Airy functions. For this purpose we approximate the bracket in
Eq.~(\ref{for_w2}), i.e., $p^2$, by a linear function of $r$ in the vicinity of
the turning points. To write this explicitly for the turning point $r_+$, we
put $p^2(r)\approx-\alpha (r-r_+)$, for which Eq.~(\ref{for_w2}) has exact
solutions $w_1(r)=\Ai(\alpha^{1/3}(r-r_+))$ and
$w_2(r)=\Bi(\alpha^{1/3}(r-r_+))$ (up to a multiplicative constant). Using the
asymptotic formulas for Airy functions~\cite{AiryFun2004} that are very accurate for
$|x|\gtrsim2$, namely
\begin{equation}
 \Ai(x)\approx\begin{cases} 
 \pi^{-1/2}(-x)^{-1/4}\sin[\frac23 (-x)^{3/2}+\pi/4] & (x<0)\\ 
 \frac12 \pi^{-1/2}x^{-1/4}\exp(-\frac23 x^{3/2}) & (x>0)\end{cases}
\label{Ai}
\end{equation} 
and
\begin{equation}
 \Bi(x)\approx\begin{cases} 
 \pi^{-1/2}(-x)^{-1/4}\cos[\frac23 (-x)^{3/2}+\pi/4] & (x<0)\\ 
 \pi^{-1/2}x^{-1/4}\exp(\frac23 x^{3/2}) & (x>0)\end{cases}
\label{Bi}
\end{equation} 
we can express the solutions $w_{1,2}(r)$ for $r<r_+$ as 
\begin{eqnarray}
 w_1(r) & \approx &
  \frac{\alpha^{1/6}}{\sqrt{\pi p}}\cos\left(\int_{r_+}^r p(r')\,\dd r'
               +\frac\pi4\right) \label{w1}\\
 w_2(r) & \approx &
  \frac{\alpha^{1/6}}{\sqrt{\pi p}}\sin\left(\int_{r_+}^r p(r')\,\dd r'
               +\frac\pi4\right)\,,
\label{w2}
\end{eqnarray} 
where the linearisation of $p^2$ was used again to write the expressions $(\pm
x)^{3/2}$ in Eqs.~(\ref{Ai}) and~(\ref{Bi}) as integrals of $p$.  Obviously,
the solutions~(\ref{w1}) and~(\ref{w2}) can be expressed as linear combinations
of the solutions~(\ref{w}) in the oscillatory region. This way Eqs.~(\ref{w1})
and~(\ref{w2}) describe the same solutions as Eq.~(\ref{w}), and a similar
consideration can be made for the evanescent region.  This allows us to express
the approximate solutions of Eq.~(\ref{for_w2}) in the evanescent and
oscillatory regions as well as in the vicinity of the turning point $r_+$ by a single formula using the Airy functions Ai and Bi. Their argument is obtained by
expressing the $x$ variable in Eqs.~(\ref{Ai}) and~(\ref{Bi}) in terms of the
integral of $p$ with the help of Eqs.~(\ref{w1}) and~(\ref{w2}). This way we
arrive at the following approximate solutions of Eq.~(\ref{for_w2}):
\begin{equation}
 f_1(r)=\Ai\left[-\sqrt[3]{\left(\frac32\int_r^{r_+}p\,\dd r\right)^{2}}\right],\quad 
 f_2(r)= \Bi\left[-\sqrt[3]{\left(\frac32\int_r^{r_+}p\,\dd r\right)^{2}}\right]\,.
\label{Airysol}
\end{equation} 
In this expression the integral is real or purely imaginary for $r<r_+$ or
$r>r_+$, respectively. Its square in these regions is positive or negative,
respectively, and so is the third root. It is easy to check that if
$p^2(r)=-\alpha (r-r_+)$, then the brackets in Eq.~(\ref{Airysol}) turn into
$\alpha^{1/3}(r-r_+)$ for all $r$, so in that case Eqs.~(\ref{Airysol})
represent the exact solutions.  Moreover, even if $p^2(r)$ is non-linear,
Eqs.~(\ref{Airysol}) still provide a very good approximation to the solutions
of Eq.~(\ref{for_w2}).  Most importantly, the positions of the nodes match very
well the positions of the nodes of the exact solution. This can be seen on the
examples in Figs.~\ref{figure-airyfun} and~\ref{figure-airyfun2}.

Eqs.~(\ref{Airysol}) enable to find the phase of the wave in the oscillatory
region very accurately. What we have to do is to find the coefficients
$c_{1,2}$ such that the superposition $w(r)=c_1f_1+c_2f_2$ satisfies the
condition $w(b)=0$. The solution is, up to a common multiplicative factor,
$c_1= \Bi(\xi)$ and $c_2=-\Ai(\xi)$, where
$\xi=-\sqrt[3]{\left(\frac32\int_b^{r_+}p\,\dd r\right)^{2}}$. Using the
asymptotic forms~(\ref{w1}), (\ref{w2}) in the oscillatory region, we find that
in this region
\begin{equation}
 w(r)\approx 
  \frac{\mbox{const}}{\sqrt{p}}\cos\left(\int_{r_+}^r p(r')\,\dd r'+\eta\right)\,,
\label{wcomb}
\end{equation} 
with the phase shift
\begin{equation}
\eta=\arctan\frac{\Bi(\xi)-\Ai(\xi)}{\Bi(\xi)+\Ai(\xi)}\,
\label{eta}
\end{equation} 
The difference between $\eta$ and $\pi/4$ gives a correction factor for
Eqs.~(\ref{k3D}) and~(\ref{k2D}) that has to be added to the curly
parentheses. A similar consideration can be made in the situation when there is
a mirror in the other evanescent region $r<r_-$. Naturally, the closer the
turning point is to the mirror, the larger will be the correction. 
This way the largest shifts will be exhibited by the levels corresponding to
the smallest values of $m$. 

\begin{figure}
\begin{center}
\begin{tabular}{ccc}
\includegraphics[width=8cm]{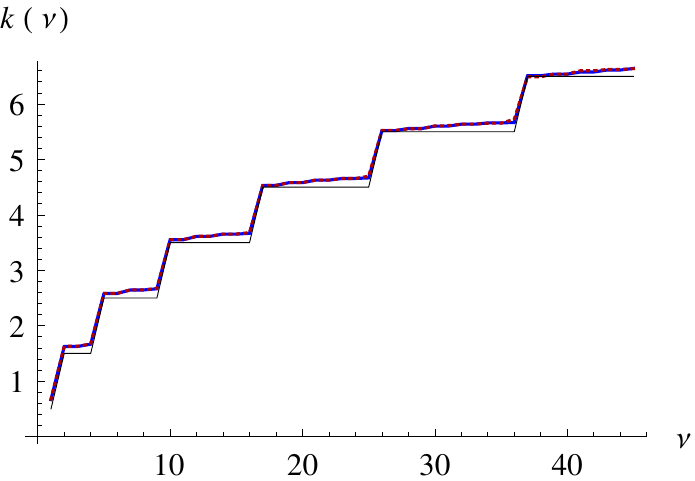}&\hspace{5mm}&
\includegraphics[width=8cm]{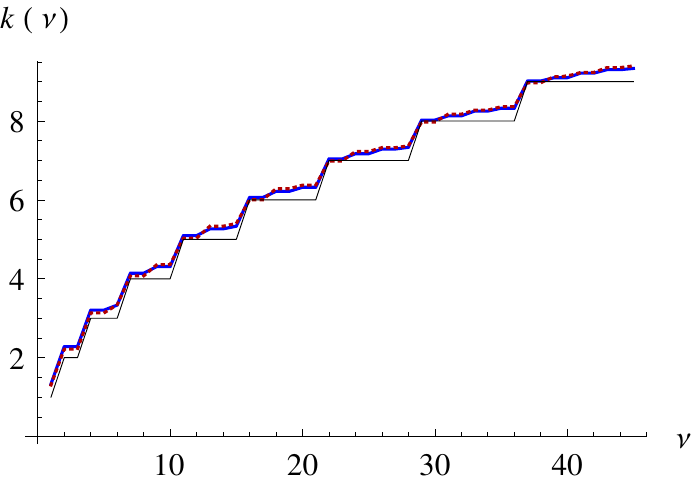}\\
(a)&&(b)
\end{tabular}
\end{center}
\caption{Comparison of the WKB spectra (blue) and numerically calculated
  spectra (dashed red) for two AIs with mirrors. (a) Kepler index profile
  surrounded by a mirror at radius $b=2$. (b) Mi\~nano profile surrounded by a
  mirror at radius $b=2$. The difference between the curves can be hardly
  seen, the relative errors make a few per cent even for the lowest levels and
  becomes completely negligible for higher frequencies. The thin black lines
  show the WKB spectra of the corresponding devices without the mirror.}
\label{fig-spectramirror}
\end{figure}

Fig.~\ref{fig-spectramirror} shows the results of our method on two examples of
AIs where the mirror has been placed at the boundary of the classically
accessible region at $r=b=2$.  The first one is the Kepler profile, the second
one is Mi\~nano lens.  The WKB spectrum is compared with the numerically
calculated spectrum.  The difference is so small that it can be hardly seen in
the pictures. For comparison, the WKB spectrum in the absence of the mirror is
also shown. The mirror obviously shifts all the levels up as could be
expected. A closer inspection of the spectrum reveals that the shift indeed
decreases with increasing $m$ due to the increasing separation of the turning
point from the mirror.

Fig.~\ref{fig-spectramirror} also clearly reveals that the spectrum is no more
so degenerate as it was in the absence of the mirror. As was shown
in~\cite{Tyc11njp}, the regularity and degeneracy of the spectrum is connected
with the quality of imaging by the device, so adding the mirror degrades
somewhat the imaging, in particular at low frequencies.  Note that this happens
even though the mirror is situated beyond the turning point $r_+$ for most rays
(i.e., for all rays with the exception of the ones with $L=0$ for which
the mirror is just at the turning point). In terms of geometrical optics,
nothing changes when the mirror is added because only the rays with $L=0$ reach
it, and even their trajectory remains unchanged. However, unlike the rays, the
wave ``feels'' the mirror even if it is located in the evanescent, i.e.,
classically inaccessible region.

\subsection{Mirror in the oscillatory region}

We will now consider the situation when the mirror is not in the evanescent,
but rather in the oscillatory region, which is the case e.g.\ of Maxwell's fish
eye mirror. Suppose that the optical medium is inside the mirror, again as in
the case of MFEM. The upper turning point $r_+$ is then located at the mirror
itself, so $w(r)$ must vanish there and so for $r<r_+$ the wave must behave
according to Eq.~(\ref{wcomb}) with $\eta=\pi/2$. This means an additional
constant phase factor of $\pi/4$ compared to Eq.~(\ref{approxcos}), which then
has to be added into the curly parentheses in Eqs.~(\ref{k3D})
and~(\ref{k2D}). Applying this result to the 2D MFEM (for which
$n(r)=2/(1+r^2)$, $\mu=2$ and $b=1$), we get for the WKB spectrum
\begin{equation}
   k_{\rm WKB}=2N+|m|+3/2\,.
\label{}
\end{equation} 
The analytical form of the spectrum is 
\begin{equation}
   k_{\rm analytic}=\sqrt{(2N+|m|+1)(2N+|m|+2)}\,.
\label{}
\end{equation} 
This way, the WKB method gives here very good results in a similar way as for
the simple Maxwell's fish eye.

\section{Conclusions}
\label{conclusions}

In this paper we have developed a method for calculating frequency spectra of
absolute instruments using the WKB approximation.  Our method is based on a the
key relation~(\ref{der}) between the optical path $S$ and the semiclassical
phase change $\Phi$ between the turning points, and on the fact that in AIs of
the first type the optical path $S$ is independent of the angular momentum.
The method proved to be very efficient for describing the spectrum accurately
even for the lowest levels. The results confirmed the previously derived
properties of the spectra, in particular that the frequencies are strongly
degenerate in AIs and that they form almost equidistantly spaced groups.  Since
these properties of the spectrum have key importance for focusing of waves by
AIs~\cite{Tyc12njp}, the WKB method shows that sharp focusing of rays in fact
implies that waves will be focused as well.  This way the WKB method provides a
nice and important bridge between geometrical and wave optics of absolute
instruments.

We applied our method also to AIs that contain mirrors; there, the Airy
functions turned out to be very helpful in describing the wave in both the
oscillatory and evanescent regions, and to calculate the eigenfrequency shifts
caused by the mirror.  When the mirror is added into the evanescent region, it
turns out that even though light rays are not influenced at all, the spectrum
can be influenced strongly. In particular, the spectrum becomes less regular,
which can somewhat degrade imaging by the absolute instrument. On the other
hand, placing the mirror in the oscillatory region leads to a constant shift in
the spectrum.

It may be somewhat surprising, but certainly very satisfactory, how accurate
results the WKB method gives for the spectra of absolute instruments. The
situation is similar to quantum mechanics where the WKB method also gives very
precise values for the energy spectrum in many situations. An interesting
question for further research could be how the WKB method can be applied to AIs
of the second type where not all points have their full images, or to even more
general devices.

\section*{Acknowledgements}

I thank Michal Lenc for his comments and acknowledge support from grant
no.\ P201/12/G028 of the Grant Agency of the Czech Republic, and from the QUEST
programme grant of the Engineering and Physical Sciences Research Council.


\begin{thebibliography}{20}

\bibitem{BornWolf} Born M and Wolf E 2006 {\em Principles of optics} (Cambridge: Cambridge University Press)
\bibitem{MFE} Maxwell J C 1854 {\em Camb. Dublin Math. J.} \textbf{8} 188
\bibitem{minano06} Mi\~{n}ano J C 2006 {\em Opt. Express} {\bf 14} 9627
\bibitem{Tyc11njp} Tyc T, Herz\'anov\'{a} L, \v Sarbort M and  Bering K 2011
{\em New J. Phys.} {\bf 13} 115004
\bibitem{Tyc11pra} Tyc T 2011 {\em Phys. Rev. A} {\bf 84} 031801(R) 
\bibitem{Ulf2009-fisheye} Leonhardt U 2009 {\em New J. Phys.} \textbf{11} 093040
\bibitem{Ulf2011} Ma Y G, Sahebdivan S, Ong C K, Tyc T and Leonhardt U 2011
  {\em New J. Phys.} {\bf 13} 033016
\bibitem{Ulf2012} Ma Y G, Sahebdivan S, Ong C K, Tyc T and Leonhardt U 2012 
{\em New J.  Phys.} {\bf 14}  025001
\bibitem{Blaikie} Blaikie R J 2010 {\em New J. Phys.} {\bf 12} 058001
\bibitem{Ulf-reply_to_Blaikie} Leonhardt U 2010 {\em New J. Phys.} {\bf 12} 058002
\bibitem{Tyc2011-nature} Tyc T and Zhang X 2011 {\em Nature} {\bf 480} 42
\bibitem{Rhiannon} Quevedo-Teruel O, Mitchell-Thomas R C and
Hao Y 2012 {\em Phys. Rev. A} {\bf 86} 053817 
\bibitem{Tyc12njp} Tyc T and Danner A 2012 {\em New J. Phys.} {\bf 14} 085023
\bibitem{Langer37} Langer R E 1934 {\em Bull. Am. Math. Soc.} {\bf 40} 545
\bibitem{Berry72} Berry M V and Mount K E 1972 {\em Rep. Prog. Phys.} {\bf 35}
  315
\bibitem{Landau} Landau L D and Lifshitz E M 1991 {\em Quantum mechanics} 
(Pergamon Press)
\bibitem{AiryFun2004} Vall\'ee O and Soares M 2004 {\em Airy functions and
  applications to physics} (London: Imperial College Press)
\end{thebibliography}
\end{document}